\documentclass[preprint,prb,aps]{revtex4}
\usepackage{amsmath,amssymb}
\usepackage{graphicx}
\usepackage{dcolumn}
\usepackage{bm}

\begin{document}
\newcommand{\ds}{\displaystyle}
\newcommand{\bea}{\begin{eqnarray}}
\newcommand{\eea}{\end{eqnarray}}

\title{A Group Theoretical Identification of Integrable Cases of the
Li\'{e}nard Type Equation $\ddot{x}+f(x)\dot{x}+g(x) = 0$ : Part I:
Equations having Non-maximal Number of Lie point Symmetries}
\author{S. N. Pandey}%
\email{snp@mnnit.ac.in (S. N. Pandey)}
\affiliation{%
Department of Physics, Motilal Nehru National Institute of
Technology, Allahabad - 211 004, India}%
\author{P. S. Bindu, M. Senthilvelan and M. Lakshmanan}%
\email{lakshman@cnld.bdu.ac.in (M. Lakshmanan)}
\affiliation{%
Centre for Nonlinear Dynamics, School of Physics, Bharathidasan
University,
 Tiruchirapalli -620 024, India
}%
\date{\today}
\begin{abstract}
We carry out a detailed Lie point symmetry group
classification of the Li\'enard type equation,
$\ddot{x}+f(x)\dot{x}+g(x) = 0$, where $f(x)$ and $g(x)$ are
arbitrary smooth functions of $x$.  We divide our analysis into
two parts.  In the present first part we isolate equations that
admit lesser parameter Lie point symmetries, namely, one, two and
three parameter symmetries, and in the second part we identify
equations that admit maximal (eight) parameter Lie-point
symmetries.  In the former case the invariant equations form a
family of integrable equations and in the latter case they form a
class of linearizable equations (under point transformations).  Further, we prove the
integrability of all of the equations obtained in the present paper
through equivalence transformations either by providing the general solution or by constructing time
independent Hamiltonians. Several of these equations are being
identified for the first time from the group theoretical analysis.
\end{abstract}

\maketitle
\section{\bf Introduction}
\label{sec1}

 In this set of two papers we perform a Lie symmetry analysis for the Li\'enard
 type equation
\begin{eqnarray}
A(x,\dot{x},\ddot{x})\equiv \ddot {x} + f(x)\dot {x} + g(x)= 0,
\label{1.1}
\end{eqnarray}
where over dot denotes differentiation with respect to time and
$f(x)$ and $g(x)$ are arbitrary smooth functions of $x$.
Notable equations from class (\ref{1.1}) include a large
number of physically important nonlinear oscillators such as the
anharmonic oscillator, force-free Helmholtz oscillator,
force-free Duffing and Duffing-van der Pol oscillators, modified
Emden-type equation (MEE), and its hierarchy, and generalized
Duffing-van der Pol oscillator equation hierarchy.  These
equations arise naturally in several physical applications.  The
outstanding representative of the class of equations (\ref{1.1})
is the modified Emden equation (also called Painlev\'e-Ince
equation), $\ddot x +\alpha x \dot x +\beta x^3 = 0$, where
$\alpha$ and $\beta$ are arbitrary parameters, which has received
considerable attention from both mathematicians and physicists for
more than a century (see for example Ref. \onlinecite{Sekar:2006cc} and references therein).

During the past three decades, immense interest has been shown
towards the search for symmetry generators of nonlinear ordinary
differential equations (ODEs) and classification of low
dimensional Lie algebras and linearization.  Eventhough Lie himself had
shown that the second order ODE of the form,
$\ddot{x}+f(t,x,\dot{x}) = 0$, can admit a maximum of eight
symmetry generators, the recent impetus came only when Wulfman and
Wybourne \cite{wulf:1976} showed that the maximal Lie group of
point transformations for the simple harmonic oscillator is eight
and the associated  group is $SL(3,R)$. Subsequently, Cervero and
Villarroel \cite{cervil:1984a} showed that the damped linear
harmonic oscillator also admits eight symmetry generators.
Thereafter, several studies were made to isolate the equations
which admit  rich Lie point symmetries by exploring their symmetry
algebras and their applications in physics and mathematics
\cite{aguirre,olver:1986,bluman:1989,Hydon:2000,cant:2002,
ibra:1999,step,hill,baum,euler,lak}.  For a recent survey of
results one may refer Ref. \onlinecite{mah1} and references therein.

Specific equations of the form (\ref{1.1}) have also been
investigated from other points of view.  For example, Noether
symmetries for certain physically important systems were also studied in
Refs. \onlinecite{cervil:1984a,lutzky,jurgen,ray}. Also
contact symmetries for the
harmonic oscillator were also explicitly constructed by Cervero
and Villarroel \cite{cervil:1984a}.  The nonlocal symmetries for
the MEE have also been studied in several papers \cite{nls1,nls2}.
Recently much interest has also been shown towards exploring
generalized symmetries, namely, $\lambda$-symmetries (also called
$c^{\infty}$-symmetries) for certain nonlinear ODEs (see for
example Ref. \onlinecite{mur1}).

The main goal of the present set of papers is to present a
detailed Lie point symmetry analysis of (\ref{1.1}) and isolate
integrable and linearizable cases explicitly.  At this point we
mention here that the study of group classification is interesting
not only from a purely mathematical point of view, but is also
important for applications.  Physical models are often constrained
with apriori requirements to symmetry properties specified by
physical laws, for example, from the Galilean or special
relativity principles.  In this work we not only isolate the
invariant equations and symmetries but also present the integrals
of motion and/or general solutions wherever possible. The
motivation for the present study comes from our recent work in
which we have considered a rather general second-order nonlinear
ODE of the form
$\ddot{x}+(k_1x^q+k_2)\dot{x}+k_3x^{2q+1}+k_4x^{q+1}+\lambda x =
0$, where $k_i$'s, $i=1,2,3,4$, $\lambda$ and $q$ are arbitrary
parameters, and identified several interesting integrable cases
and for certain equations we have also derived explicit solutions
of both oscillatory and non-oscillatory types\cite{Sekar:2006c}.
Interestingly, we demonstrated that a system which is very close
to the MEE, namely $\ddot{x}+\alpha
x\dot{x}+\frac{\alpha^2}{9}x^3+\lambda x = 0$, possesses certain
basic properties which is very uncommon to nonlinear
oscillators\cite{vkc1:2005}: It is a conserved Hamiltonian system (of
nonstandard type)
admitting amplitude independent harmonic oscillations.   While the
above mentioned equations are specific examples of (1), the
question arises as to whether there exist other
integrable/linearizable second order ODEs belonging to this class.
In this and the accompanying paper (referred to as II), we present
a systematic analysis towards this goal.  In this way we classify
both integrable and linearizable equations which belong to the
Li\'enard type system (1).

Even though the Lie's algorithm is in principle a straightforward
one, the group classification for the present problem is reduced
to integration of a complicated overdetermined system of partial
differential equations for the infinitesimal symmetry functions
for arbitrary forms of  $f(x)$ and $g(x)$ in (1). While solving
these determining equations we have to choose all the symmetry
functions (see Eqs. (\ref{2.11}) and (\ref{2.12}) below) not equal to
zero in order to obtain the maximal Lie point symmetries.  On the
other hand considering the special case of one or more of the symmetry
functions to be equal to zero also, we obtain a spectrum of integrable
equations. In this way we are able to classify (i) systems with
non-maximal symmetries (less than eight) (ii) systems with maximal
symmetries (eight).  Note that for second order ODEs the dimension of
the Lie vector space cannot be four, five, six and seven.  In the present
paper we focus our attention
only on the equations associated with lesser parameter Lie-point
symmetry groups, that is one, two and three parameter Lie-point
symmetry groups. However, within this classification, we identify
a wider class of important integrable equations which are
invariant under 2-parameter point symmetry group. Many of them are being
identified for the first time from the group theoretical analysis.
In the second paper we identify all the equations admitting
maximal number of symmetries which also turn out to be
linearizable.  The forms of $f$ and $g$ which lead to only two
symmetry generators are as follows:
\begin{subequations}
\begin{align}
\mbox{(i)~~} & f  =  k_2+f_1x^{q}, &&  g  =
\frac{k_2^2(q+1)}{(q+2)^2}x +\frac{k_2f_1}{(q+2)}x^{q+1}
+g_1x^{2q+1},\\
\mbox{(ii)~~} & f  =  -f_1+\lambda_2 \mbox{log}(x), && g  =
g_1 x -(\frac{\lambda_2f_1}{2}+\frac{\lambda_2^2}{4})x\log x +
\frac{\lambda_2^2}{4}x(\log x)^2,\\
\mbox{(iii)~~} & f  =  \frac{f_1}{x^2}, &&
g = -\frac{A^2}{4}x+\frac{Af_1}{2x}
+\frac{g_1}{x^3}, \\
\mbox{(iv)~~} & f  =  \frac{\lambda_1}{\lambda_2}
+f_1e^{-\lambda_2x}, && g  =
-\frac{\lambda_1}{\lambda_2^2}f_1e^{-\lambda_2x}
+g_1e^{-2\lambda_2x}-\frac{\lambda_1^2}{\lambda_2^3}. \label{s1}
\end{align}
\end{subequations}
(Here $f_1,g_1,k_2,\lambda_1,\lambda_2,q,A$ are all constants).
All the above equations are pointed out to be integrable through equivalence
transformations.  We also
identify that the only system which admits a three parameter symmetry
group is the Pinney-Ermakov equation where $f=0$ and $g=\omega^2 x
-\frac{\tilde{g}}{x^3}$ ($\omega,\tilde{g}$: constants).

It is well known that the infinitesimal generators of a given Lie group form a Lie algebra.  The Lie
algebras constituted by Lie vector fields are widely used in the
integration of differential equations\cite{olver:1986}, group classification
of ODEs and PDES\cite{Horwath}, in geometric control theory and in the theory
of systems with superposition priniciples\cite{Carinena} and in different
schemes for numerical solution of differential equations\cite{Bourlioux}.
A vast amount of works is available in the literature on the classification of
realizations of finite dimensional Lie algebras on the real and complex planes.
For example, the realizations of all possible complex Lie algebras of dimensions
no greater than four were listed by Lie himself\cite{Lie}.  Recently
Gonzalez-Lopez et al have provided the Lie's classification of realizations of
complex Lie algebras\cite{Lopez} and extended it to the real case.  A complete
set of inequivalent realizations of real Lie algebras of dimension no greater than
than four in the vector fields on a space of an arbitrary (finite) number of variables
was constructed in Ref. \onlinecite{Poyovych}.  For
more details on the classification of Lie vector fields one may refer the recent work
of Ref. \onlinecite{Nesterenko} and references therein.  On the other hand, in our paper we focuss our attention on constructing Lie vector fields for the class of equations resulting out of Eq. (1) alone and discuss
their integrability.

 The plan of the paper is as follows.
In the following section, we present the Lie's algorithm for Eq.
(1) and discuss the solvability of the determining equations. A
careful analysis of our investigations show that one should
consider two separate cases, namely (i) the symmetry function
$(i)\; b = 0$ and (ii)\;$b \neq 0$, while solving the determining
equations. Since the former case admits three symmetry
functions, $a(t,x), c(t,x), d(t,x)$, while classifying the
integrable equations, we consider the possibilities $(i)\;d=0, a,c
\neq 0$, and $(ii)\;c=0, a,d \neq 0$ separately and bring out the
equations that are invariant under both the possibilities in Sec.
III.  We also discuss sub-cases in both the cases $(i)$ and
$(ii)$. Further, we prove that the system (1) does not admit a
three parameter Lie point symmetry group when both $f(x),g(x) \neq
0$. In Sec. IV we investigate the equivalence transformations for Eq. (1)
and show that they lead to integrable forms.  In Sec. V we consider the
special case in which either $f(x)$
or $g(x)$ is equal to zero and identify the associated integrable
equations in this class. The notable example in this class
includes Pinney-Ermakov equation. In Appendix A, we present some
details on the Hamiltonian structure of an integrable equation
that arises in the case $(i)$.  In Appendix B, we point out briefly
some notable equations that are included in the most general
equation (corresponding to (2a)).  In Appendix C, we discuss the
method of solving the integrable
equation identified as integrable in this paper corresponding to (2b).  The
Liouville
integrability of the two other integrable equations identified in the category
$c=0, a,d \neq 0$ are presented in Appendices D and E.  Finally, we present our
conclusions in Sec. VI.

\section{\bf Determining Equations for the Infinitesimal Symmetries}
\label{sec2} We consider the one dimensional nonlinear Li\'enard
type system of the form (\ref{1.1}). Let the evolution equation
be invariant under the one parameter Lie group of infinitesimal
transformations
\begin{eqnarray}
&&\tilde{t}  =� t+\epsilon \xi(t,x)+O(\epsilon^2), \nonumber\\
&&\tilde{x}  =  x+\epsilon \eta(t,x)+O(\epsilon^2), \quad \epsilon \ll 1,
\label{2.1}
\end{eqnarray}
where $\xi$ and $\eta$ represents the infinitesimal symmetries associated
with the variables $t$ and $x$ respectively. The associated infinitesimal
generator can be written as
\begin{eqnarray}
X  =�\xi(t,x)\frac{\partial}{\partial t}+ \eta(t,x)\frac{\partial}
{\partial x}.
\label{2.2}
\end{eqnarray}
Eq. (\ref{1.1}) is invariant under the action of (\ref{2.2}) iff
\begin{eqnarray}
X^{(2)}(A)|_{A=0} = 0,
\label{2.3}
\end{eqnarray}
where
\begin{eqnarray}
X^{(2)} = \xi\frac{\partial }{\partial t}+ \eta\frac{\partial }{\partial x}+
\eta^{(1)}\frac{\partial }{\partial\dot{x}}+
\eta^{(2)}\frac{\partial }{\partial\ddot{x}}
\label{2.4}
\end{eqnarray}
is the second prolongation \cite{olver:1986,bluman:1989} in which
\begin{eqnarray}
\eta^{(1)} = \dot{\eta}-\dot{x}\dot{\xi},\;\;\;
\eta^{(2)} = \ddot{\eta}-\dot{x}\ddot{\xi}-2\ddot{x}\dot{\xi},
\label{2.5}
\end{eqnarray}
and dot denotes total differentiation.
By analysing Eq. (\ref{2.3}) we get the following determining equations:
\begin{eqnarray}
&&\xi_{xx}  =  0,
\label{2.7} \\
&&\eta_{xx}-2\xi_{tx}+2f\xi_x  =  0,
\label{2.8} \\
&&2\eta_{tx}-\xi_{tt}+f\xi_t+3g\xi_x+\eta f_x  = 0,
\label{2.9} \\
&&\eta_{tt}-(\eta_x-2\xi_{t})g+f\eta_t+\eta g_x  =  0,
\label{2.10}
\end{eqnarray}
where subscripts denote partial derivatives.

Solving Eqs. (\ref{2.7}) and (\ref{2.8}) we obtain
\begin{eqnarray}
\xi = a(t)+b(t)x
\label{2.11}
\end{eqnarray}
and
\begin{eqnarray}
\eta = \dot{b}{x}^2-2b\Im(x) +c(t)x+d(t),
\label{2.12}
\end{eqnarray}
where
\begin{eqnarray}
\Im_x = F(x) = \int_0^{x}f(x')dx'\;\;\;\; \mbox{and}\;\;\;\;
\Im_{xx}=f(x),
\label{2.13}
\end{eqnarray}
and $a(t)$, $b(t)$, $c(t)$ and $d(t)$ are arbitrary functions of $t$. With
these forms of
$\xi$ and $\eta$, Eqs. (\ref{2.9}) and (\ref{2.10}) can be rewritten as
\begin{eqnarray}
(\dot{b} x^2-2b\Im+cx+d)f_x+(\dot{a}+\dot{b}x)f+3bg-4\dot{b}F+
3\ddot{b}x+2\dot{c}-\ddot{a}=0,
\label{2.14}
\end{eqnarray}
and
\begin{eqnarray}
&&(\dot{b}x^2-2b\Im+cx+d)g_x-(c-2\dot{a}-2bF)g - 2\ddot{b}\Im
\nonumber\\
&&\qquad\qquad\qquad\qquad+(\ddot{b} x^2-2\dot{b}\Im + \dot{c}x +\dot{d})f
+\dddot{b}x^2+\ddot{c}x +\ddot{d}  = 0,
\label{2.15}
\end{eqnarray}
respectively.
Solving Eqs. (\ref{2.14}) and (\ref{2.15}) for the given forms of $f(x)$ and $g(x) $we can get the
infinitesimal symmetries.

The foremost and simplest solution for (\ref{2.14})
and (\ref{2.15}) for any form of $f$ and $g$ is $a=\mbox {constant},
b,c,d =0$.
In other words, one immediately gets the time translation generator
$X = \frac{\partial}{\partial t}$ irrespective of the form of $f$ and $g$.
Our motivation here is to find explicit forms
of $f$ and $g$ which admit more number of symmetries. For this purpose we
solve Eqs. (\ref{2.14}) and (\ref{2.15}) in the following way. Rewriting
Eq. (\ref{2.14}), we get
\begin{eqnarray}
g=\frac{1}{3b} [-(\dot{b}x^2-2b\Im+cx+d)f_x-(\dot{a}+\dot{b}x)f
+4\dot{b}F-3\ddot{b}x-2\dot{c}+\ddot{a} ], \; b \neq 0.
\label{2.16}
\end{eqnarray}
Thus the existence of Lie point symmetries of the general form
(\ref{2.11}) and (\ref{2.12}) with $b \neq 0$ introduces an
interrelation between the functions $f$ and $g$.  However, this relation
has to be compatible with the second determining equation (\ref{2.15}).  Thus using
(\ref{2.16}) into (\ref{2.15}), one can obtain an equation for $f$,
(which also involves all the four symmetry functions)
which fixes its form as well as  the associated symmetries
systematically.  This is carried out in the following paper II, where
we show explicitly that maximal (eight) number of Lie point symmetries
exists only for the case $f_{xx} = 0$ and for $f_{xx} \neq 0$ necessarily
requires the symmetry function $b=0$.  Consequently, one has to consider the case $b=0$ in
Eqs. (\ref{2.14}) - (\ref{2.15})  separately because of the
condition $ b \neq 0$ in Eq. (\ref{2.16}).  Thus it is of interest
to consider two separate cases associated with Eqs.
(\ref{2.14})-(\ref{2.15}):

Case (i) $\;b = 0:$ Since we assume one of the symmetry
functions to be zero the determining equations lead us to lesser Lie
point symmetries alone (one, two and three symmetries).

Case (ii) $\;b \neq 0:$ In this case we solve the full determining
equations which in turn lead us to the maximal (eight) Lie-point
symmetry group (as well as other lesser point symmetries while $ b
\neq 0$) when $f_{xx} = 0$.

In this paper, we analyze in detail only the Case (i) and present
the results of the other case in the subsequent paper II.

\subsection{\bf Alternate Way}
One may also note here that one can proceed in an alternate way to analyze the
determining equations (\ref{2.14}) and (\ref{2.15}) for compatibility to
determine the forms of $f(x)$ and $g(x)$ and the associated symmetries.
For example, Eq. (\ref{2.14}) can be rewritten as
\begin{eqnarray}
\ddot{a} = (\dot{b} x^2-2b\Im+cx+d)f_x+(\dot{a}+\dot{b}x)f+3bg-4\dot{b}F+
3\ddot{b}x+2\dot{c}.
\label{new1}
\end{eqnarray}
Differentiating the above equation with respect to $x$ and rearranging one can express $\ddot{b}$
in terms of $\dot{a}, \dot{b}, c$ and $d$.  Differentiating again the resultant equation
with respect to $x$ and simplifying the latter we find
\begin{eqnarray}
\dot{a}f_{xx} & = & -[(\dot{b}x^2-2b\Im+cx+d)_{xx}f_x-2(\dot{b}x^2-2b\Im+cx+d)_xf_{xx}
\notag\\
& & \quad (\dot{b}x^2-2b\Im+cx+d)f_{xxx}-\dot{b}xf_{xx}-3bg_{xx}+2\dot{b}f_x.
\label{new2}
\end{eqnarray}
Similarly from Eq. (\ref{2.15}) one can obtain expressions for $\ddot{c}$ and
$\ddot{d}$ and finally arrive at an expression of the form
\begin{eqnarray}
0 & = & -[(\dot{b}x^2-2b\Im+cx+d)g_x-(c-2\dot{a}-2bF)g - 2\ddot{b}\Im
\nonumber\\
& & +(\ddot{b} x^2-2\dot{b}\Im + \dot{c}x +\dot{d})f
+\dddot{b}x^2]_{xx}.
\label{new3}
\end{eqnarray}
To find a compatible solution of Eqs. (\ref{new2}) and (\ref{new3}) one may substitute
for $\dot{a}$ from (\ref{new2}) into (\ref{new3}) and analyze the resultant equation to  find
the allowed forms of $f$ and $g$ and the associated symmetries.  However, in practice we find the
method leads to very lengthy and laborious calculations.  On the other hand in the procedure we adopt we find that
for the case $f_{xx} \neq 0$, one necessarily requires $b=0$, see Sec. V in the following paper II.
Consequently the analysis given in Sec. III follows naturally.

On the other hand, for the case $f_{xx}=0$, Eq. (\ref{new2}) leads to the condition
\begin{eqnarray}
0  =  2bff_{x}-3bg_{xx}.
\label{new4}
\end{eqnarray}
which is consistent with (\ref{new3}).  From (\ref{new4}) one can immediately write
$f=f_1+f_2x$ and $g=\frac{1}{3}f_1f_2x^2+\frac{1}{9}f_2^2x^3+g_1x+g_2$.
To fix the corresponding symmetries one has to resubstitute the forms of $f$ and $g$
in the original determining equations and solve them consistently.  This is carried out in
Paper II.

\section{\bf Lie symmetries of Li\'enard type systems -
Lesser parameter symmetries: Case $\mathbf{b=0}$}
We now explore the nature of the evolution
equations which possess lesser parameter Lie point symmetries.
Considering Eqs. (\ref{2.14}) and (\ref{2.15}), we now assume the
function $b=0$ to obtain the following determining equations,
\begin{eqnarray}
(cx+d)f_x+\dot a f+2\dot c-\ddot a=0
\label{4.1}
\end{eqnarray}
and
\begin{eqnarray}
(cx+d)g_x-(c-2\dot{a})g+(\dot{c}x+\dot d)f+\ddot c x+\ddot d=0,
\label{4.2}
\end{eqnarray}
respectively.
We note here that the determining equation (\ref{4.1}) for the
function $f$  does not involve the function $g$. As a
consequence an explicit form for $f$ can be determined by direct integration.
Now substituting this form of $f$ into
equation (\ref{4.2}) we can derive
the corresponding form of $g$.

It is a well known fact \cite{olver:1986,bluman:1989} that a
second order ODE  admits only $1,2,3$ or $8$ parameter Lie point
symmetries (which we will see explicitly for Eq. (1) also in the
present as well as in the follow up paper II). In Sec.2 we noted
that the most general equation which is invariant under the one
parameter Lie point symmetry group is the general equation
(\ref{1.1}) itself with arbitrary form of $f(x)$ and $g(x)$ since
there is no explicit appearance of $t$ in the equation and the
associated symmetry generator is $\frac{\partial}{\partial t}$.
But for $b=0$, we explicitly show in the following that only two
parameter symmetries exit when both $f \neq 0$ and $g \neq 0$ and
obtain their specific forms, while three parameter symmetries can
also exist only when $f = 0, g \neq 0 $.  Finally, in the case $f
\neq 0,\; g=0$, the second order ODE (1) can be rewritten as a
first order equation which in turn can be integrated by
quadratures straightforwardly.  So we do not investigate this last
category in this work.

\subsection{\bf 2-parameter Lie point symmetries}
Rewriting Eq.~(\ref{4.1}), we have
\begin{eqnarray}
f_x + \frac{\dot{a}}{c(x+\frac{d}{c})}f =
\frac{\ddot{a}-2\dot{c}}{c(x+\frac{d}{c})}. \label{ps02}
\end{eqnarray}
Since $f$ should be a function of $x$ alone (vide Eq. (1)), we choose
\begin{eqnarray}
\frac{\dot{a}}{c} = \lambda_1,\;\;\frac{\ddot{a}-2\dot{c}}{c} = \lambda_2,\;\;
\frac{d}{c} = \lambda_3,
\label{ps03a}
\end{eqnarray}
where $\lambda_1,\;\lambda_2$ and $\lambda_3$ are constants.  Then
Eq. (\ref{ps02}) becomes
\begin{eqnarray}
f_x + \frac{\lambda_1}{(x+\lambda_3)}f = \frac{\lambda_2}{(x+\lambda_3)}.
\label{ps04a}
\end{eqnarray}
Integrating Eq. (\ref{ps04a}) one obtains
\begin{eqnarray}
f = \frac{\lambda_2}{\lambda_1}+f_1(x+\lambda_3)^{-\lambda_1} ,
\label{ps05a}
\end{eqnarray}
where $f_1$ is also an arbitrary constant.  Note that since $f_1, \lambda_1,\;\lambda_2$
and $\lambda_3$ now occur in the form for $f(x)$ which determine the ODE (1), they are not
symmetry parameters but rather they are the {\it system parameters}.

 One can proceed further by constructing the
associated form of $g$  by solving Eq. (\ref{4.2}) and classify the
invariant
 equations. To start with, let us first classify for convenience the equations which are
 invariant
under 2-parameter Lie-point symmetry group. To deduce these
equations, we consider the two possibilities  $(i)~ d = 0; \; a, c
\neq 0$, $\;(ii)~c= 0; \; a, d \neq 0$ and also the sub-cases in
both of them. We further note that due to the fact the system
always admits translational symmetry, from Eqs. (\ref{1.1}) and
(\ref{2.11}), it is clear that $a$ cannot be zero.  So we need not
consider the case $a=0$, $c,d \neq 0$.  On the other hand if both
$c$ and $d$ are simultaneously zero, then $g = 0$ as may be
inferred from Eq. (\ref{4.2}). Finally, at the end of this section
we consider the cases where none of the functions $a,b,c$ are zero
and show that even here only $2$-parameter symmetries exist.

\subsubsection{\bf Case 1 $\mathbf{d=0;\;
a,\;c \neq 0 \; (\lambda_1 \neq 0, \lambda_2 \neq 0)} $}
The choice $d(t)=0$ with $a(t),\; c(t)\neq 0$
leads us to several interesting new integrable equations, as we
see below. Solving Eq. (\ref{ps03a}), with $d(t) =0$ and so
$\lambda_3 = 0$,  one can
obtain explicit forms for the functions $a$ and $c$ as
\begin{eqnarray}
a = a_1+\frac{\lambda_1}{\lambda_2}(\lambda_1-2)c_1
e^{(\frac{\lambda_2}{\lambda_1-2})t},\quad
c=c_1e^{(\frac{\lambda_2}{\lambda_1-2})t},
\label{ps14}
\end{eqnarray}
where $a_1$ and $c_1$ are two arbitrary (symmetry) parameters, which lead to a two
parameter Lie-point symmetry group.  Substituting
Eqs. (\ref{ps05a}) and (\ref{ps14}) into (\ref{4.2}) with $d=0$, we
get
\begin{eqnarray}
g_x+\frac{(2\lambda_1-1)}{x}g=\frac{-2\lambda_2^2(\lambda_1-1)}
{\lambda_1(\lambda_1-2)^2}-\frac{\lambda_2f_1}{(\lambda_1-2)}
x^{-\lambda_1}.
\label{ps25}
\end{eqnarray}
Integrating Eq.~(\ref{ps25}), we obtain
\begin{eqnarray}
g=\frac{\lambda_2^2(1-\lambda_1)}{\lambda_1^2(\lambda_1-2)^2}x
+\frac{\lambda_2f_1}{(2-\lambda_1)\lambda_1}x^{1-\lambda_1}+g_1
x^{1-2\lambda_1},
\label{ps26}
\end{eqnarray}
where $g_1$ is another integration constant. The above forms of $f$ and
$g$ (vide Eqs.~(\ref{ps05a}) and (\ref{ps26}), respectively) fix Eq.
(\ref{1.1}) to the form
\begin{eqnarray}
\ddot{x}+\bigg(\frac{\lambda_2}{\lambda_1}+f_1x^{-\lambda_1}\bigg)
\dot{x}
+\frac{\lambda_2^2(1-\lambda_1)}{\lambda_1^2(\lambda_1-2)^2}x
+\frac{\lambda_2f_1}{(2-\lambda_1)\lambda_1}x^{1-\lambda_1}
+g_1x^{1-2\lambda_1}=0. \label{ps27}
\end{eqnarray}
For the sake of neatness, we rewrite $\lambda_1=-q, \; q \ne 0$,
and $\lambda_2=-k_2q$, where $k_2$ is an arbitrary parameter, in
the above equation so that we obtain
\begin{eqnarray}
\ddot{x}+\bigg(k_2+f_1x^{q}\bigg)\dot{x}
+\frac{k_2^2(q+1)}{(q+2)^2}x
+\frac{k_2f_1}{(q+2)}x^{q+1}
+g_1x^{2q+1}=0,\label{ps29}
\end{eqnarray}
where $k_2$, $f_1$ and  $q$ are nothing but system parameters.
Eq.~(\ref{ps29}) is the most general equation that is invariant under the two
parameter Lie point symmetry group with infinitesimal symmetries
\begin{eqnarray}
\xi=a_1-\frac{1}{k_2} (q+2)c_1 e^{\ds\frac{k_2q}{(q+2)} t},\quad
\eta=c_1e^{\ds\frac{k_2 q}{(q+2)}t} x.
\label{ps27aa}
\end{eqnarray}
The corresponding  infinitesimal generators read
\begin{eqnarray}
X_1=\frac{\partial}{\partial t},\quad
X_2=e^{\ds\frac{k_2 q}{(q+2)}t}
\bigg[-\frac{(q+2)}{k_2}\frac{\partial}{\partial t}
+x\frac{\partial}{\partial x}\bigg].\label{ps27a}
\end{eqnarray}
The commutation relation between the vector fields $X_1$ and $X_2$ is given
by
\begin{eqnarray}
[X_1,X_2] = \frac{k_2 qX_2}{(q+2)}.\label{ps27b}
\end{eqnarray}

\subsubsection{\bf Integrability of Eq. (\ref{ps29}) for arbitrary
values of $q$}

Eq. (\ref{ps29}) is the most general integrable equation which is
invariant under the two parameter symmetry group (\ref{ps27aa}).
Now we discuss the integrability of (\ref{ps29}) briefly here. By
introducing the transformation
\begin{align}
w = xe^{\frac{k_2}{(q+2)}t},\;\;\; z =
-\frac{(q+2)}{qk_2}e^{-\frac{qk_2}{(q+2)}t} \label{ps34aa}
\end{align}
where $w$ and $z$ are new dependent and independent variables,
respectively, one can transform (\ref{ps29}) to the form
\begin{eqnarray}
w''+\alpha w^qw'+\beta w^{2q+1}=0, \label{ps34b}
\end{eqnarray}
where $\alpha = \frac{(q+2)^2f_1}{2k_2^2q^2}$ and $\beta =
\frac{(q+2)4g_1}{4k_2^4q_2^4}$. Eq. (\ref{ps34b}) has been analyzed
from different
perspectives. For example,  Lemmer and Leach
\cite{nls1} have studied the hidden symmetries of Eq.
(\ref{ps34b}).  Feix et al. \cite{feix} have shown that through a
direct transformation to a third order equation the above
Eq.~(\ref{ps34b}) can be integrated to obtain the general solution
for the specific choice of the parameter ${\beta}$, namely
${\beta}=\frac{\alpha}{(q+2)^2}$. For this choice of ${\beta}$, the
general
solution of (\ref{ps34b}) can be written as
\begin{eqnarray}
x(t)=\bigg(\frac{(2+3q+l^2)(t+I_1)^l}{l(t+I_1)^{q+1}
+(2+3q+q^2)I_2}\bigg)^{\frac{1}{q}}, \;\; I_1,
I_2:\;\mbox{arbitrary constants}. \label {hie05}
\end{eqnarray}
For the same parametric choice recently we have shown that this
equation can be linearized to a free particle equation through a
generalized linearizing transformation so that the solution of the
nonlinear equation can be constructed from the solution of the
linearized equation \cite{Sekar:2006a}. However, our very recent
studies show that Eq. (\ref{ps34b}) admits time independent
Hamiltonian description for all values of $\alpha$ and $\beta$.  By
introducing appropriate canonical transformation to the Hamilton's
canonical equation of motion one can integrate the resultant
equations straightforwardly and obtain the general solution (for
more details one may see Refs. \onlinecite{Sekar:2006c,Gvkc1}).  For
convenience, the Hamiltonian structure of the above equation is
indicated in the Appendix A.  We also point out briefly other
notable equations included in (\ref{ps29}) in Appendix B.

\subsubsection{\bf Integrable equations with $d=0, \; a,c \neq 0$ ($\lambda_1 =0 \; (q=0),
\lambda_2 \neq 0$)}

Earlier while deriving the form (\ref{ps05a}) for $f(x)$ we assumed that $\lambda_1 \neq 0$.
Now let us consider the case  $\lambda_1=0$. From Eq. (\ref{ps03a}) we
find that in this case
\begin{align}
a=a_1, \hskip 5 pt c=c_1 \mbox {e}^{\frac{-\lambda_2t}{2}},
\label{se1}
\end{align}
where $a_1$ and $c_1$ are two arbitrary symmetry parameters which again
lead us to a two parameter symmetry group. Solving Eq. (\ref{ps02}), with the above forms
of $a$ and $c$,
we obtain
\begin{align}
f(x) = -f_1+\lambda_2 \mbox{log}(x), \label{se2}
\end{align}
where $f_1$ is an integration constant. Substituting Eq. (\ref{se2})
into Eq. (\ref{4.2}) with $d =0$, we get
\begin{align}
g_x - \frac{g}{x} + \frac{\lambda_2}{2}(f_1-\lambda_2 \log x)
+\frac{\lambda_2^2}{4} = 0. \label{se3}
\end{align}
Integrating Eq. (\ref{se3}), we obtain the following specific form
for $g$,
\begin{align}
 g  =  g_1 x-(\frac{\lambda_2f_1}{2}+\frac{\lambda_2^2}{4})x\log
x + \frac{\lambda_2^2}{4}x(\log x)^2,
\label{se4}
\end{align}
where $g_1$ is an integration constant. Using Eqs. (\ref{se2})
and (\ref{se4}) in Eq. (\ref{1.1}), we have the following
nonlinear ODE,
\begin{align}
 \ddot x + (-f_1+\lambda_2 \mbox{log}(x))\dot x + g_1 x
-(\frac{\lambda_2f_1}{2}+\frac{\lambda_2^2}{4})x\log x +
\frac{\lambda_2^2}{4}x(\log x)^2 =0,
\label{se5}
\end{align}
which is invariant under the following infinitesimal symmetries
\begin{align}
\xi = a_1, \hskip 5pt \eta = c_1\mbox{e}^{\frac{-\lambda_2t}{2}} x.
\label{se6}
\end{align}
The associated symmetry generators take the form
\begin{align}
X_1=\frac{\partial}{\partial t}, \hskip 5 pt X_2 =
\mbox{e}^{\frac{\lambda_2t}{2}} x \frac{\partial}{\partial x}.
\label{se7}
\end{align}
The integrability of Eq. (\ref{se5}) can be proved straightforwardly which we
indicate in Appendix C.

\subsubsection{\bf Integrable equations with $d=0, a,c \neq 0$
($\lambda_1  \neq 0, \lambda_2=0) \;\; $}
While deriving (\ref{ps05a}) we assumed that $\lambda_2 \neq 0 $. Now
we analyse the case $\lambda_2 = 0$ with $\lambda_1 \neq 0 $. In
this case we find that the compatible solution exists for either
$\lambda_1 \neq 2$ or $\lambda_1 = 2$. In the first case by
repeating the previous analysis we find that $f = f_1x^{-\lambda_1}$ and $g =
g_1 x^{(1-2\lambda_1)}$, where $f_1$ and $g_1$ are two arbitrary
parameters so that Eq.~(1) becomes
\begin{align}
\ddot{x}+f_1x^{-\lambda_1}\dot{x}+g_1 x^{(1-2\lambda_1)} = 0.
\label{ss1}
\end{align}
The associated infinitesimal generators turn out to be
\begin{align}
X_1=\frac{\partial}{\partial t}, \hskip 5 pt X_2 =
t\frac{\partial}{\partial t} + \frac{x}{\lambda_1}
\frac{\partial}{\partial x}.
\label{ss2}
\end{align}
Eq. (\ref{ss1}) exactly coincides with (\ref{ps34b}) by redefining $\lambda_1 = -q$,
and so the integrability of (\ref{ss1}) can be extracted from (\ref{ps34b}).

In the second case, namely, $\lambda_1 = 2$, we obtain that the
following form of equation for (1),
\begin{align}
\ddot{x}+\frac{f_1}{x^2}\dot{x}-\frac{A^2}{4}x+\frac{Af_1}{2x}
+\frac{g_1}{x^3} = 0, \label{ss3}
\end{align}
where $A$ is an arbitrary parameter, which is invariant under the
two parameter infinitesimal symmetry generators,
\begin{align}
X_1=\frac{\partial}{\partial t}, \hskip 5 pt X_2 =e^{-At}\left(
\frac{-2}{A}\frac{\partial}{\partial t} +x
\frac{\partial}{\partial x}\right). \label{ss4}
\end{align}
We discuss the integrability of Eq. (\ref{ss4}) in Appendix D.

In this and previous sub-sections we discussed the cases $(i)\; \lambda_1=0,
\lambda_2 \neq 0$ and $(ii)\;
\lambda_2=0, \lambda_1 \neq 0$.  Finally, for the third case, namely $(iii)\;
 \lambda_1=0,
\lambda_2 = 0$, one
gets $a=constant = a_1,\; c=constant = c_1$.  The invariant equation turns out
to be
the linear damped harmonic oscillator equation
$\ddot{x}+f_1\dot{x}+g_1x=0$, where $f_1$ and $g_1$ are arbitrary parameters.  The
associated infinitesimal vector fields are $X_1 = \frac{\partial}{\partial t},\;\;
X_2 = x\frac{\partial}{\partial x}$.  It is known that damped harmonic oscillator
equation admits eight parameter symmetry group.  Since one of the symmetry
functions is zero we obtained only a two parameter symmetry group.  The full
symmetry group of the damped harmonic oscillator will be discussed in paper II.

\subsubsection{\bf Case 2$\;\;\;\;\mathbf{ c=0, \;\; a,\;d\neq0}$:
Integrable equation}
In the previous sub-section we considered the case $d=0,
a,c \neq 0$. Now we focuss our attention on the case $c=0, a,d
\neq 0$ and fix the forms of $f$ and $g$ which are invariant under
the corresponding symmetry transformations. Restricting to $c=0$
and $a,d \neq 0$ in (\ref{4.1}), we have
\begin{eqnarray}
f_x+\frac{\dot{a}}{d}f= \frac{\ddot{a}}{d}.
\label{ps03}
\end{eqnarray}
As before, since $f$ has to be a function of $x$ alone,
we choose
\begin{eqnarray}
\frac{\ddot{a}}{d}=\mbox{constant}=\lambda_1, \quad
\frac{\dot{a}}{d}=\lambda_2=\mbox{constant},
 \label{ps05}
\end{eqnarray}
so that Eq. (\ref{ps03}) becomes
\begin{eqnarray}
f_x+\lambda_2 f = \lambda_1.
\label{ps05b}
\end{eqnarray}
Solving (\ref{ps05}) we  obtain
\begin{eqnarray}
a =
a_1+\frac{\lambda_2^2}{\lambda_1}d_1e^{\frac{\lambda_1}{\lambda_2}t},
\quad d = d_1e^{\frac{\lambda_1}{\lambda_2}t}
, \label{ps06}
\end{eqnarray}
where $a_1$ and $d_1$ are two integration constants which are also
the two symmetry parameters.  Integration of Eq. (\ref{ps05b})
leads us to
\begin{eqnarray}
f = \frac{\lambda_1}{\lambda_2}+f_1e^{-\lambda_2 x}, \label{ps07}
\end{eqnarray}
where $f_1$ is an arbitrary constant.
Substituting (\ref{ps07}) into (\ref{4.2}), with $c=0$, we get
\begin{eqnarray}
g_x+2\lambda_2g+\frac{\lambda_1}{\lambda_2}f_1e^{-\lambda_2x}
+\frac{2\lambda_1^2}{\lambda_2^2} = 0.
\label{ps08}
\end{eqnarray}
Integrating (\ref{ps08}) we obtain
\begin{eqnarray}
g =-\frac{\lambda_1}{\lambda_2^2}f_1e^{-\lambda_2x}
+g_1e^{-2\lambda_2x}-\frac{\lambda_1^2}{\lambda_2^3}, \label{ps09}
\end{eqnarray}
where $g_1$ is an integration constant. Eqs.~(\ref{ps07}) and
(\ref{ps09}) fix the equation (\ref{1.1}) to the specific form
\begin{eqnarray}
\ddot{x}+\bigg[\frac{\lambda_1}{\lambda_2} +f_1e^{-\lambda_2x}\bigg]\dot{x}
-\frac{\lambda_1}{\lambda_2^2}f_1e^{-\lambda_2x}
+g_1e^{-2\lambda_2x}-\frac{\lambda_1^2}{\lambda_2^3}=0.\label{ps010}
\end{eqnarray}

We note that in the above $\lambda_1$, $\lambda_2$, $g_1$, and
$f_1$ are system parameters. Eq. (\ref{ps010}) is invariant under
the following two parameter Lie point symmetries
\begin{eqnarray}
\xi = a(t) = a_1+\frac{\lambda_2^2}{\lambda_1}
d_1e^{\frac{\lambda_1}{\lambda_2}t},\;\;
\eta = d_1e^{\frac{\lambda_1}{\lambda_2}t},\label{ps011}
\end{eqnarray}
where $d_1$ and $a_1$ are the symmetry parameters. The associated
infinitesimal generators are
\begin{eqnarray}
X_1 = \frac{\partial}{\partial t}, \qquad
X_2 = e^{\frac{\lambda_1}{\lambda_2}t}\bigg[\frac{\lambda_2^2}{\lambda_1}
\frac{\partial}{\partial t}+\frac{\partial}{\partial x}
\bigg].\label{ps012}
\end{eqnarray}

The integrability of the Eq. (\ref{ps010}) is discussed in
Appendix E.
\subsection{\bf Non-existence of 3-parameter symmetry group in the
general case $a,c,d \neq 0$}
Finally, we consider the general case in which none of the
functions $a,\;c$ and $d$ are  zero.  In this case, the function
$f$ takes the form given in Eq. (\ref{ps05a}).  The functions
$a,c$ and $d$ can be fixed by solving the Eq. (\ref{ps03a}).
Doing so we find
\begin{eqnarray}
a=a_1+\frac{\lambda_1(\lambda_1-2)}{\lambda_2}c_1e^{\frac{\lambda_2}
{(\lambda_1-2)}t},\quad
c=c_1e^{\frac{\lambda_2}{(\lambda_1-2)}t},\quad
d=\lambda_3c_1e^{\frac{\lambda_2}{(\lambda_1-2)}t},\label{ps52}
\end{eqnarray}
where only $a_1$ and $c_1$ are the symmetry parameters.  So in this case also only
2-parameter symmetries exist and no 3-parameter symmetry group is possible.

Substituting  the forms $f,a,c$ and $d$ into Eq. (\ref{4.2}) and
simplifying the resultant equation one obtains
\begin{eqnarray}
g_x+\frac{(2\lambda_1-1)}{(x+\lambda_3)}g=\frac{2\lambda_2^2(1-\lambda_1)}
{\lambda_1(\lambda_1-2)^2}+\frac{\lambda_2f_1}
{(2-\lambda_1)(x+\lambda_3)^{\lambda_1}},\label{ps53}
\end{eqnarray}
where $g_1$ is an integration constant.  One can directly integrate
(\ref{ps53}) to obtain
\begin{eqnarray}
g=\bigg(\frac{\lambda_2}{\lambda_1}\bigg)^2\frac{(1-\lambda_1)}
{(2-\lambda_1)^2}
(x+\lambda_3)+\bigg(\frac{\lambda_2}{\lambda_1}\bigg)
\frac{f_1(x+\lambda_3)^{(1-\lambda_1)}}{(2-\lambda_1)}
+g_1(x+\lambda_3)^{(1-2\lambda_1)}. \label{ps54}
\end{eqnarray}
Inserting the forms (\ref{ps05a}) and (\ref{ps54}) in (\ref{1.1})
we get
\begin{eqnarray}
&&\ddot{x}+\bigg(\frac{\lambda_2}{\lambda_1}
+\frac{f_1}{(x+\lambda_3)^{\lambda_1}}\bigg)\dot{x}
+\bigg(\frac{\lambda_2}{\lambda_1}\bigg)^2\frac{(1-\lambda)}{(2-\lambda_1)^2}
(x+\lambda_3)\nonumber\\
&&\qquad\qquad\qquad+\bigg(\frac{\lambda_2}{\lambda_1}\bigg)
\frac{f_1(x+\lambda_3)^{(1-\lambda_1)}}{(2-\lambda_1)}
+g_1(x+\lambda_3)^{(1-2\lambda_1)}=0.\label{ps55}
\end{eqnarray}

It is interesting to note that the system possesses only two
parameter Lie point symmetries. The infinitesimal symmetries and
generators are
\begin{eqnarray}
\xi =
a_1+\frac{\lambda_1(\lambda_1-2)}{\lambda_2}c_1e^{\frac{\lambda_2}
{(\lambda_1-2)}t},\quad
\eta=c_1e^{\frac{\lambda_2}{(\lambda_1-2)}t}
(x+\lambda_3),\label{ps52a}
\end{eqnarray}
and
\begin{eqnarray}
X_1=\frac{\partial}{\partial t}, \quad
X_2=e^{\frac{\lambda_2}{(\lambda_1-2)}t}\left(\frac{\lambda_1(\lambda_1-2)}
{\lambda_2} \frac{\partial}{\partial t} + (x+\lambda_3)
\frac{\partial}{\partial x}\right)
\end{eqnarray}
respectively.

Redefining $X=x+\lambda_3$ in (\ref{ps55}), the resultant equation
coincides exactly with the integrable Eq. (\ref{ps27}).  The symmetry
generators also coincide with the ones given in Eq.
(\ref{ps27aa}). So effectively no new nonlinear ODE is identified
even when all the three symmetry functions are simultaneously
nonzero.

Thus we conclude that the system (1) does not admit a three
parameter Lie-point symmetry group when $f(x), g(x) \neq 0$ while
the symmetry function $b(x) = 0$ in
(\ref{2.7})-(\ref{2.10}).  Further, the only equations which admit two
parameter symmetry group alone are the four nonlinear ODEs given by
Eqs. (\ref{ps29}),(\ref{se5}),(\ref{ss3}) and (\ref{ps010}).

\section{Equivalence transformations}
We have shown in the above section that the identified evolution equations, namely
Eqs. (\ref{ps29}), (\ref{se5}), (\ref{ss3}) and (\ref{ps010}), admitting two parameter Lie
point symmetries
can be transformed into integrable equations
(\ref{ps29}), (\ref{ja21}), (\ref{D1}) and (\ref{ka31})  respectively,
through appropriate transformations.  In this section we give a group
theoretical interpretation for these results through equivalence transformations (ETs).
We invoke the equivalence transformations
and give an explanation for the results since
the group classification problem is closely related to the concept of equivalence of
equations of the above forms with respect to transformations, see for example Ref. \onlinecite{Ovsyan}.

Considering our original differential equation (1),
let us consider a set of smooth, locally one-to-one transformations $\mathcal{T}: (t,x,f,g) \longrightarrow
(T,X,f_1,g_1) $ of the space $\mathbb{R}^4$ that act by the formulas
\begin{align}
T=F(t,x),\;\; X=G(t,x), \;\; f_1=H(t,x,f),\;\; g_1=L(t,x,g)
\label{eqv1}
\end{align}
A transformation is called an Equivalence Transformation (ET) of the equality $\ddot{x}=-f(x)\dot{x}
-g(x)$ if it transforms the equation
\begin{align}
\ddot{x}=-f(x)\dot{x}-g(x)
\label{eqv2}
\end{align}
to an equation of the same form
\begin{align}
\ddot{X}=-f_1(X)\dot{X}-g_1(X).
\label{eqv3}
\end{align}
In this case, Eqs. (\ref{eqv2}) and (\ref{eqv3}) and the functions $\{f(x)$,$g(x)\}$ and
$\{f_1(X)$,$g_1(X)\}$ are equivalent\cite{Ovsyan}.

It is a proven fact that equivalent equations admit similar groups (for local transformations)
and ET is a similarity
transformation.  That is, if (\ref{eqv2}) admits the group $E$ then (\ref{eqv3}) also admits
a group similar to it for local transformations.

Substituting the transformation (\ref{eqv1}) into Eq. (\ref{eqv3}) we get
\begin{align}
& (G_t^2F_tf_1+G_t^3g_1)+\dot{x}[(G_t^2F_x+2G_tF_tG_x)f_1+3G_t^2G_xg_1]
+ \dot{x}^2[3G_x^2G_tg_1
\nonumber\\
& \qquad +(F_tG_x^2+2G_tF_xG_x)f_1]+
\dot{x}^3\left[F_xG_x^2f_1+G_x^3g_1\right]= -(G_t+\dot{x}G_x)[(F_{tt}
\nonumber\\
 & \qquad
 +2\dot{x}F_{tx}+\dot{x}^2F_{xx}
-F_x(\dot{x}f+g)]+(F_t+\dot{x}F_x)[(G_{tt}+2\dot{x}G_{tx}
\nonumber\\
& \qquad +\dot{x}^2G_{xx}-G_x(\dot{x}f+g)],
\label{eqv4}
\end{align}
where the subscripts denote partial derivative with respect to that variable.
Equating the coefficients of different powers of $\dot{x}^n,\;n=0,1,2,3$, we get
\begin{align}
 F_xG_x^2f_1+G_x^3g_1 & =  F_xG_{xx}-G_xF_{xx},
\label{eqv5a}\\
 (F_tG_x^2+2G_tF_xG_x)f_1]+3G_x^2G_tg_1 & =  F_tG_{xx}+2F_xG_{tx}-G_tF_{xx}-2G_xF_{tx},
\label{eqv5b}\\
(G_t^2F_x+2G_tF_tG_x)f_1+3G_t^2G_xg_1 & =
 -2G_tF_{tx}-G_xF_{tt}+fF_xG_t+2F_tG_{tx}\nonumber\\
  & \qquad \qquad  -F_tG_{x}-F_xG_{tt},
\label{eqv5c}\\
 G_t^2F_tf_1+G_t^3g_1 & =  -G_tF_{tt}+gF_{x}G_{t}+F_tG_{tt}-gG_{x}F_{t}.
\label{eqv5d}
\end{align}
Solving Eqs. (\ref{eqv5a}) and (\ref{eqv5b}) consistently
we find $G_x = 0$ and $F_{xx} = 0$.  As a result
one gets
\begin{eqnarray}
G = \alpha(t), \qquad
F = \beta(t)x+\gamma(t),
\label{eqv7b}
\end{eqnarray}
where $\alpha, \beta$ and $\gamma$ are arbitrary functions of $t$.
Substituting Eq. (\ref{eqv7b}) in (\ref{eqv5c}) and (\ref{eqv5d}) and simplifying the resultant equations we get
\begin{align}
\dot{\alpha}^2\beta f_1 & =  f\beta \dot{\alpha}+\beta\ddot{\alpha}-2\dot{\alpha}
\dot{\beta},
\label{eqv8a}\\
\dot{\alpha}^2(\dot{\beta}x+\dot{\gamma})f_1+\dot{\alpha}^3g_1
& =  -\dot{\alpha}(\ddot{\beta}x+\ddot{\gamma})+g\beta\dot{\alpha}
+\ddot{\alpha}(\ddot{\beta}x+\ddot{\gamma}).
\label{eqv8b}
\end{align}
From Eq. (\ref{eqv8a}) we can obtain an expression which connects the transformed function $f_1$
with the original function $f$ of the form
\begin{eqnarray}
f_1 = \frac{f}{\dot{\alpha}}+\frac{\ddot{\alpha}}{\dot{\alpha}^2}
-\frac{2\dot{\beta}}{\beta\dot{\alpha}}.
\label{eqv9}
\end{eqnarray}
Substituting (\ref{eqv9}) in (\ref{eqv8b}) and simplifying the resultant equation we arrive at
\begin{eqnarray}
g_1 = \frac{\beta g}{\dot{\alpha}^2}-\frac{(\dot{\beta}x+\dot{\gamma})}{\dot{\alpha}^2}f
+\frac{2\dot{\beta}(\dot{\beta}x+\dot{\gamma})}{\beta\dot{\alpha}^2}
-\frac{(\ddot{\beta}x+\ddot{\gamma})}{\dot{\alpha}^2}.
\label{eqv9a}
\end{eqnarray}
Thus we obtain the general ET
\begin{eqnarray}
T=\alpha(t),\;\; X = \beta(t)x+\gamma(t),\;\;
f_1 = \frac{f}{\dot{\alpha}}+\frac{\ddot{\alpha}}{\dot{\alpha}^2}
-\frac{2\dot{\beta}}{\beta\dot{\alpha}},
\nonumber\\
g_1 = g_1 = \frac{\beta g}{\dot{\alpha}^2}-\frac{(\dot{\beta}x+\dot{\gamma})}{\dot{\alpha}^2}f
+\frac{2\dot{\beta}(\dot{\beta}x+\dot{\gamma})}{\beta\dot{\alpha}^2}
-\frac{(\ddot{\beta}x+\ddot{\gamma})}{\dot{\alpha}^2}.
\label{eqv9aa}
\end{eqnarray}
Since we have already identified only four equations (vide Eqs. (\ref{ps29}), (\ref{se5}), (\ref{ss3}) and (\ref{ps010})) that are invariant under two parameter Lie
point symmetries within the class of equations (1) we consider only these four equations and present
our result. Now solving Eqs. (\ref{eqv9aa}) with the given form of $f$ and $g$ one obtains the following
result
\begin{align}
& \text {Case 1 (Eq.(\ref{ps29}))} \;\;\;
\alpha = -\frac{(q+2)}{qk_2}e^{-\frac{qk_2}{(q+2)}}t,\quad
\beta = e^{-\frac{k_2}{(q+2)}}t, \quad \gamma = 0
\nonumber\\
& \qquad \qquad \qquad \qquad \text{so that} \;\; f_1 = \alpha X^q \;\;\; g_1 = \beta X^{2q+1}
\nonumber\\
& \text {Case 2 (Eq.(\ref{se5}))} \;\;\;
\alpha = t,\quad
\beta = e^{-\frac{\lambda_2}{2}\int log[x(t)]dt}, \quad \gamma = 0
\nonumber\\
& \qquad \qquad \qquad \qquad \text{so that} \;\; f_1 = {\text constant} \;\;\; g_1 = X
\nonumber\\
& \text {Case 3 (Eq.(\ref{ss3}))} \;\;\;
\alpha = \frac{1}{A}e^{At},\quad \beta = e^{At}, \quad \gamma = 0
 \nonumber\\
& \qquad \qquad \qquad \qquad \text{so that} \;\; f_1 = \frac{1}{X^2} \;\;\; g_1 = \frac{1}{X^3}
\nonumber\\
& \text {Case 4 (Eq.(\ref{ps010}))} \;\;\;
\alpha = -\frac{\lambda_2}{\lambda_1}e^{-\frac{\lambda_1}{\lambda_2}},\quad
\beta = 1, \quad \gamma = -\frac{\lambda_1}{\lambda_2^2}t
\nonumber\\
& \qquad \qquad \qquad \qquad \text{so that} \;\; f_1 = e^{\lambda_2U} \;\;\;
g_1 = e^{-2\lambda_2U}
\end{align}
It directly follows that with the above form of $f_1$ and $g_1$, Eq. (\ref{eqv3}) takes the
form of (\ref{ps29}), (\ref{ja21}), (\ref{D1}) and (\ref{ka31})) respectively, which were shown to be integrable.
\section{\bf Lie Symmetries of Eq. (\ref{1.1}) with $f(x)=0$ or $g(x) = 0$}
Next we consider the special case of Eq. (\ref{1.1}) with
$f(x)=0$, that is,
\begin{align}
\ddot x +g(x) =0. \label{se8}
 \end{align}
In the following we focuss our attention only on the case $ b=0$
so that we have $ \xi = a(t) , \hskip 5 pt \eta = c(t) x +d(t)$.
Eqs. (\ref{2.9}) and (\ref{2.10}) with $b(x)=0$ and $f(x)=0$ give
rise to the following conditions, respectively,
\begin{align}
\ddot a - 2\dot c =0 \label{se14},
\end{align} and
\begin{align}
g_x+\Big(\frac{2 \dot a -c}{c x + d}\Big) g+\frac{\ddot c x +\ddot
d} {c x + d}=0.
\label{se15}
\end{align}
Since $g(x)$ should be a function of $x$ alone, we choose
\begin{eqnarray}
\frac{2\dot a}{c}-1 = \lambda_1,\;\;
\frac{d}{c} = \lambda_2,\;\;
\frac{\ddot c}{c} = -\lambda_3,
\label{se16}
\end{eqnarray}
where $\lambda_1,\lambda_2$ and $\lambda_3$ are constant parameters. Note that
the above implies $\displaystyle{\frac{\ddot{d}}{c}}=\lambda_2\displaystyle{\frac{\ddot{c}}{c}}
=\lambda_2\lambda_3$.

Solving (\ref{se16}) we find that the solution exists either for the parametric
choice $\lambda_1 \neq 3$ or $\lambda_1 = 3$.
The respective infinitesimal symmetries are
\begin{align}
& \xi = a(t) = a_1+a_2 t,\;\; \eta = c(t)x+d(t) = \frac{2a_2x}{(1+\lambda_1)}
+\frac{2\lambda_2a_2}{(1+\lambda_1)}, \;\;\;\;\;\; \lambda_1 \neq
3
 \label{31a}\\
& \xi = a(t) = a_1 -\frac{(1+\lambda_1)}{2\sqrt{\lambda_3}}(c_2\cos \sqrt{\lambda_3}t
-c_1\sin \sqrt{\lambda_3}t), \nonumber\\
& \eta = c(t)x+d(t) = a_1-(c_1\cos \sqrt{\lambda_3}t+c_2\sin \sqrt{\lambda_3}t)(
x+\lambda_2),\quad \quad \;\; \lambda_1 = 3 \label{se15ab}
\end{align}
The respective invariant equations turn out to be
\begin{align}
\ddot{x}+\frac{g_1}{(x + \lambda_2)^{\lambda_1}} = 0, \qquad
\qquad \qquad \qquad \lambda_1 \neq 3, \label{31a1}
\end{align}
and
\begin{align}
\ddot{x}+\frac{\lambda_3}{4}(x+\lambda_2)+
\frac{g_1}{(x+\lambda_2)^3} = 0, \qquad \;\; \lambda_1 = 3.
\label{se15aa}
\end{align}
Thus Eq. (\ref{31a1}) admits a two parameter symmetry group with the
generators
\begin{align}
X_1 = \frac{\partial}{\partial t},\;\;\;
X_2 = t\frac{\partial}{\partial t}+\frac{2(x+\lambda_2)}{(1+\lambda_1)}
\frac{\partial}{\partial x}.
\label{31a2}
\end{align}
On the other hand Eq. (\ref{se15aa}) admits a three parameter symmetry group with
the symmetry generators
\begin{align}
& X_1 = \frac{\partial}{\partial t},\;\;\;
X_2 = \sin \sqrt{\lambda_3}t \left(\frac{(1+\lambda_1)}{2\sqrt{\lambda_3}}
\frac{\partial}{\partial t}
+(x+\lambda_2)\frac{\partial}{\partial x}\right),\nonumber\\
& X_3 = \cos \sqrt{\lambda_3}t \left(\frac{(1+\lambda_1)}{2\sqrt{\lambda_3}}
\frac{\partial}{\partial t}
+(x+\lambda_2)\frac{\partial}{\partial x}\right).
\label{31a2}
\end{align}
Redefining $x+\lambda_2 = X$ in Eqs. (\ref{31a1}) and (\ref{se15aa}) we get
\begin{align}
\ddot{X}+\frac{g_1}{X^{\lambda_1}} = 0, \qquad \qquad \qquad
\qquad \qquad \qquad \qquad
\lambda_1 \neq 3.
\label{31a3}\\
\ddot{X}+\omega^2X- \frac{\tilde{g}}{X^3} = 0, \quad \omega^2 =
\frac{\lambda_3}{4}, \;\;\; \tilde{g} = -g_1, \quad \; \lambda_1 =
3, \label{se16aa}
\end{align}
Equation in (\ref{31a3}) corresponds to a conservative
Hamiltonian system ($ H =
\frac{p^2}{2}+\frac{g_1}{(1-\lambda_1)}X^{1-\lambda_1}$) and so
the Liouville integrability is assured.  On the other hand Eq. (\ref{se16aa})
is nothing but the Pinney-Ermakov
equation, whose origin, properties and the method of finding its
general solution have been discussed widely in the contemporary
nonlinear dynamics literature (see for example Ref. \onlinecite{carine} and
references therein). For the sake of completeness we give the
general solution of this equation as
\begin{align}
X = \frac{1}{A\omega}\sqrt{(\omega^2A^4-\tilde{g})\sin^2(\omega
t+\phi)+\tilde{g}},\;\; \label{se16ab}
\end{align}
It has also been shown that Eq. (\ref{se16aa}) can be transformed to harmonic
oscillator equation through suitable nonlocal transformation and from the
solution of the latter one can construct the solution for the nonlinear
equation.  For more details one may refer \cite{Sekar:2006a}.

Finally, for $g(x)=0$, Eq. (\ref{1.1}) can be written as \bea
\ddot x +f(x)\dot x =0. \label{se35} \eea Eq. (\ref{se35}) can be
transformed to a first order equation by a trivial change of
variable which in turn can also be integrated trivially.
So we do not discuss the symmetries of this equation here.

\section{Conclusions}

In the present paper we have investigated the Li\'{e}nard type equation (1)
in the framework of modern
group analysis of differential equations.  Even though the integrability
properties of some of the specific equations coming under
the Li\'{e}nard type have been discussed in the
literature, we have identified all those equations which admit only two and
three parameter symmetry groups.

To identify the integrable equations belonging to the class (1) we
have deduced all the equations that are invariant under one, two and
three parameter Lie point symmetries. Obviously the general Eq.
(\ref{1.1}) does not contain the variable $t$ explicitly and so it always
admits a time translational generator.  However, we have
demonstrated that several equations admit two parameter Lie point
symmetry groups.  In particular these equations correspond to four
specific forms of the functions $f(x)$ and $g(x)$ in (1), see Eq. (2), namely
Eqs. (\ref{ps29}), (\ref{se5}), (\ref{ss3}) and (\ref{ps010}).
These equations have been deduced here through a group theoretical
point of view alone.  We have also discussed the integrability
properties of these equations briefly and shown the existence of equivalence
transformations. After analyzing the
Lie point symmetries we have also shown that Li\'{e}nard type
equation does not admit a three parameter symmetry group when both
$f(x),g(x) \neq 0$ in Eq. (1).  However, in the sub-case,
$f(x) = 0$, one can find that the well known Pinney-Ermakov equation
is the only equation which is invariant under a three parameter Lie point
symmetry group.

In this paper we have restricted our attention only on the
non-maximal Lie point symmetry groups. The question which
naturally arises is what happens if one considers the more general
case, $b \neq 0$, vide Eq. (\ref{2.16}). Such an analysis allows
us to isolate a class of equations admitting eight
parameter symmetries. We will present the results in the follow-up
paper II.
\section*{Acknowledgments}
One of us (SNP) is grateful to the Centre for Nonlinear Dynamics,
Bharathidasan University, Tiruchirappalli, for warm
hospitality.
 The work of SNP forms part of a
Department of Science and Technology, Government of India
sponsored research project. The work of MS forms part of a
research project sponsored by National Board for Higher
Mathematics, Government of India.  The work of ML forms part of a
Department of Science and Technology (DST), Ramanna Fellowship and is
also supported by a DST-IRHPA research project.

\appendix
In the following, we briefly discuss the integrability properties of
the equations derived in Sec. III. To begin with let us consider the
Liouville integrability of Eqs. (33).

\section{Time independent Hamiltonian for (\ref{ps34b})}

Recently, we have studied the integrability of (\ref{ps29}) or
equivalently (\ref{ps34b}) and found that it admits time
independent integrals for all values of the parameters $\alpha$ and
$\beta$\cite{Sekar:2006c,Gvkc1}.  From the time independent
integrals we have identified the following time independent
Hamiltonian for (\ref{ps34b}), namely,
\begin{eqnarray}
H=\left\{\label{emden_ham}
\begin{array}{ll}
\frac{(r-1)}{(r-2)}p^{\frac{(r-2)}{(r-1)}}
-\frac{(r-1)}{r} \hat{\alpha} pw^{q+1}, & \alpha^2>4\beta (q+1)\\
\frac{\hat{\alpha}}{2}pw^{q+1}+\log(\frac{1}{p}),&\alpha^2=4\beta (q+1)\\
\frac{1}{2}\log\left[\frac{w^{2(q+1)}}{(q+1)^2}
\sec^2[\frac{\omega}{(q+1)}w^{q+1}p]\right]-\frac{\hat{\alpha}}{2}pw^{q+1},&\alpha^2<4\beta (q+1),
\end{array}
\right.
\label{A1}
\end{eqnarray}
\vskip 4pt
\noindent where the corresponding canonically conjugate momentum is defined by
\begin{eqnarray}
p=\left\{
\begin{array}{ll}
\left(\dot{w}+\frac{(r-1)}{r}
\hat{\alpha}w^{q+1}\right)^{(1-r)}, \qquad \qquad \qquad \qquad \quad \alpha^2 \ge 4\beta (q+1)\\
\frac{(q+1)}{\omega w^{q+1}}\tan^{-1}\left[\frac{\alpha w^{q+1}
+2(q+1)\dot{w}}{2\omega w^{q+1}}\right]  \qquad \qquad \qquad \quad \; \alpha^2 < 4\beta (q+1),
\end{array}
\right. \label{A2}
\end{eqnarray}
where $r = \frac{\alpha}{2\beta(q+1)}(\alpha\pm\sqrt{\alpha^2-4\beta(q+1)}$,
$\omega=\frac{1}{2}\sqrt{4\beta(q+1)-\alpha^2}$ and $\hat{\alpha} = \frac{\alpha}{q+1}$.
For more details about the derivation of the above Hamiltonian one
may refer to Ref. \onlinecite{Gvkc1}.   The time independent
Hamiltonian ensures the Liouville integrability of (\ref{ps29}) or
(\ref{ps34b}).

\section{Notable integrable equations in (\ref{ps29})}

Besides the general case, $q = arbitrary$, Eq. (\ref{ps29})
encompasses several known integrable equations of contemporary
interest.  The interesting equations can be identified by
appropriately choosing the parameter $q$ as we demonstrate briefly
in the following.

For example, choosing $q=1$ in (\ref{ps29}) one gets the generalized
MEE,
\begin{eqnarray}
\ddot{x}+(k_2+f_1x)\dot{x}+\frac{2k_2^2}{9}x+\frac{k_2f_1}{3}x^2+g_1x^3
= 0, \label{B1}
\end{eqnarray}
Eq. (\ref{B1}) can be transformed into the MEE, $w''+f_1ww'+g_1w^3
= 0$, by introducing a transformation $w= xe^{\frac{k_2}{3}t}$ and
$z = -\frac{3}{k_2}e^{\frac{-k_2}{3}t}$.  The Hamiltonian
structure for this equation can be extracted from (\ref{A1}) by
restricting $q=1$ in the latter relations. The restriction $f_1
=0$  in (\ref{B1}) provides us the force-free Duffing oscillator whose
invariance and integrability properties have been discussed in
Refs.\onlinecite{lak,Parthasarathy:1990}.  With the choice
$f_1=3,g_1=1,k_2=0$, the resultant equation becomes a linearizable
one whose invariance and integrability properties have been discussed in
detail in Refs. \onlinecite{mahomed:85,mahomed:89,duarte:87a,leach:88}.

The case $q=2$ in (\ref{ps29}) gives us
\begin{eqnarray}
\ddot{x}+(k_2+f_1x^2)\dot{x}+\frac{3k_2^2}{16}x+\frac{k_2f_1}{4}x^3+g_1x^5
= 0. \label{B2}
\end{eqnarray}
The
explicit form of the Hamiltonian can be fixed from (\ref{A1}) by
restricting $q=2$ in the latter relations.  We note here that Eq. (\ref{B2})
also includes several known integrable equations. The notable examples are
force-free Duffing-van der Pol oscillator equation ($g_1=0$) and the second
equation in the MEE hierarchy ($f_1=0,g_1=\frac{1}{16}$).

Finally, we note that one may also recover specific equations like
the force-free Helmholtz oscillator and the associated Lie
symmetries can be obtained by appropriately choosing the value of
the parameter $q$.  Choosing $q=\frac{1}{2}$ and $f_1=0$ in
(\ref{ps29}) one gets the force-free Helmhotz oscillator.  The
symmetries of (\ref{ps27aa}) with $q = \frac{1}{2}$ coincide exactly
with the one reported in Ref. \onlinecite{Almendral:1995}.

\section{Method of integrating Eq. (\ref{se5})}
The solution of Eq. (\ref{se5}) can be constructed from the
solution of the damped harmonic oscillator using a general
procedure given by us sometime ago in Ref. \onlinecite{Sekar:2006a}. For
example, let us consider a linear ODE of the form
\begin{align}
\ddot{U}+\alpha\dot{U}+g_1U = 0,
\label{ja21}
\end{align}
where $\alpha$ and $g_1$ are arbitrary parameters.
By introducing a nonlocal transformation of the form
$U = xe^{\frac{\lambda_2}{2}\int^t \log x(t')dt'}$ in the linear ODE (\ref{ja21})
the latter can be brought to the form
\begin{align}
 \ddot x + (\alpha+\frac{\lambda_2}{2}+\lambda_2\mbox{log}(x))\dot x + g_1 x
+\frac{\alpha\lambda_2}{2}x\log x +
\frac{\lambda_2^2}{4}x(\log x)^2 =0.
\label{ja22}
\end{align}
Now redefining the constants $(\alpha+\frac{\lambda_2}{2})=-f_1$
in (\ref{ja22}) one exactly ends up with (\ref{se5}).

Following the procedure given in Ref. \onlinecite{Sekar:2006a} one can obtain the
general solution for (\ref{ja22}) from the linear equation.

\section{Method of integrating Eq. (\ref{ss3})}
Eq. (\ref{ss3}) can be transformed to the equation of the form
\begin{align}
\ddot{U}+\frac{f_1}{U^2}\dot{U}+\frac{g_1}{U^3} = 0, \label{D1}
\end{align}
through the transformation $U=xe^{\frac{A}{2}t},\;\; Z
=\frac{1}{A}e^{A t}$.  Eq. (\ref{D1}) admits
Hamiltonian structure for all values of $f_1$ and $g_1$.  The
underlying Hamiltonian reads
\begin{eqnarray}
H=\left\{\label{ham}
\begin{array}{ll}
\frac{(r-1)}{(r-2)}p^{\frac{(r-2)}{(r-1)}}
+\frac{(r-1)f_1}{r}\frac{p}{U}, & \qquad f_1^2 >-4g_1\\
\log(\frac{1}{p})-\frac{f_1}{2}\frac{p}{U},& \qquad f_1^2 = -4g_1\\
\frac{f_1}{2}\frac{p}{U}+\frac{1}{2}
\log\left[\frac{1}{U^2}\sec^2[\frac{-\omega
p}{U}\right], & \qquad f_1^2<-4g_1,
\end{array}
\right. \label{D2}
\end{eqnarray}
where the canonical conjugate momentum is defined by
\begin{eqnarray}
p=\left\{
\begin{array}{ll}
\frac{1}{(r-1)}(\dot{U}-\frac{(r-1)f_1}{rU})^{(1-r)}, & \qquad f_1^2\ge4g_1\\
\frac{-U}{\omega}\tan^{-1}[\frac{f_1-2U\dot{U}}{2\omega}], & \qquad f_1^2<4g_1
\end{array}
\right. \label{A2}
\end{eqnarray}
where $r=\frac{-f_1}{2g_1}(f_1\pm\sqrt{f_1^2+4g_1})$ and
$\omega=\frac{1}{2}\sqrt{-4g_1-f_1^2}$.

The time independent Hamiltonian given above ensures the Liouville
integrability of Eq. (\ref{D1}).

\section{Method of integrating Eq. (\ref{ps010})}
By introducing a transformation $U = x -\frac{\lambda_1}{\lambda_2^2}t$ and
$z = \frac{-\lambda_2}{\lambda_1}e^{-\frac{\lambda_1}{\lambda_2}t}$ in
(\ref{ps010}) the latter can be transformed into the form
\begin{align}
U''+f_1e^{\lambda_2U}U'+g_1e^{-2\lambda_2U}
 = 0. \quad \quad ('=\frac{d}{dz})
\label{ka31}
\end{align}
Eq. (\ref{ka31}) can be rewritten in the form
form
\begin{align}
U''+f_1f(U)U'+\tilde{g}_1f(U)\int f(U)dU = 0,
\label{ka32}
\end{align}
where $f(U)=e^{-\lambda_2U}$ and $\tilde{g}_1=-\lambda_2 g_1$.
Eq. (\ref{ka32}) admits time independent Hamiltonian for all values of $f_1$ and
$g_1$.  The respective Hamiltonians are
\begin{eqnarray}
H=\left\{\label{ham}
\begin{array}{ll}
\frac{(r-1)}{(r-2)}p^{\frac{(r-2)}{(r-1)}}
+\frac{(r-1)f_1}{r\lambda_2}pe^{-\lambda_2U}, & \;\;f_1^2 > 4\tilde{g}_1\\
\log[p]+\frac{f_1p}{2\lambda_2}e^{-\lambda_2U},& \;\; f_1 = 4\tilde{g}_1\\
\frac{1}{2}\log\left[\frac{e^{-2\lambda_2U}}{\lambda_2^2}
\sec^2[\frac{\omega p e^{-\lambda_2U}}{-\lambda_2}]\right]+\frac{f_1}{2\lambda_2}pe^{-\lambda_2U}
,& \;\; f_1 < 4\tilde{g}_1,
\end{array}
\right. \label{E1}
\end{eqnarray}
where the canonically conjugate momentum is defined by
\begin{eqnarray}
p=\left\{
\begin{array}{ll}
{[U'+\frac{(1-r)}{r}\frac{f_1e^{-\lambda_2U}}{\lambda_2}]^{1-r}}, & \qquad f_1^2\ge 4\tilde{g_1} \\
\frac{-\lambda_2 e^{\lambda_2U}}{\omega}\tan^{-1}
[\frac{2\lambda_2U'-f_1e^{-\lambda_2U}}{2\omega e^{-\lambda_2U}}], & \qquad f_1^2 < 4\tilde{g_1}
\end{array}
\right. \label{A2}
\end{eqnarray}
where $r=\frac{f_1}{2g_1}(f_1\pm\sqrt{f_1^2-4\tilde{g}_1}),\;\;
\omega=\frac{1}{2}\sqrt{4\tilde{g}_1-f_1^2}$.

One may note that the above Hamiltonian resembles the Hamiltonian structure of (\ref{A1}).
The reason for this is that both the Hamiltonians (\ref{E1}) and (\ref{A1}) can be generated
from the time independent Hamiltonian of the damped harmonic oscillator by suitable
nonlocal transformation.  For more details about this nonlocal transformation
one may refer to Ref. \onlinecite{Gvkc1}.

\end{document}